\documentclass[a4paper,10pt]{article}   
\usepackage{osajnl2} 
\usepackage{xspace}
\usepackage{graphicx}
\usepackage{tabularx}

\newcommand{\ignore}[1]{}
\providecommand{\ao}{}
\renewcommand{\ao}{adaptive optics (AO)\renewcommand{\ao}{AO\xspace}\renewcommand{\Ao}{AO\xspace}\xspace}
\newcommand{\Ao}{Adaptive optics (AO)\renewcommand{\ao}{AO\xspace}\renewcommand{\Ao}{AO\xspace}\xspace}
\newcommand{\wfs}{wavefront sensor (WFS)\renewcommand{\wfs}{WFS\xspace}\renewcommand{\wfss}{WFSs\xspace}\xspace}
\newcommand{\wfss}{wavefront sensors (WFSs)\renewcommand{\wfs}{WFS\xspace}\renewcommand{\wfss}{WFSs\xspace}\xspace}
\newcommand{\shwfs}{Shack-Hartmann \wfs (SHWFS)\renewcommand{\shwfs}{SHWFS\xspace}\xspace}
\newcommand{\dm}{deformable mirror (DM)\renewcommand{\dm}{DM\xspace}\renewcommand{\dms}{DMs\xspace}\renewcommand{\Dms}{DMs\xspace}\renewcommand{\Dm}{DM\xspace}\xspace}
\newcommand{\dms}{deformable mirrors (DMs)\renewcommand{\dm}{DM\xspace}\renewcommand{\dms}{DMs\xspace}\renewcommand{\Dms}{DMs\xspace}\renewcommand{\Dm}{DM\xspace}\xspace}
\newcommand{\Dms}{Deformable mirrors (DMs)\renewcommand{\dm}{DM\xspace}\renewcommand{\dms}{DMs\xspace}\renewcommand{\Dms}{DMs\xspace}\renewcommand{\Dm}{DM\xspace}\xspace}
\newcommand{\Dm}{Deformable mirror (DM)\renewcommand{\dm}{DM\xspace}\renewcommand{\dms}{DMs\xspace}\renewcommand{\Dms}{DMs\xspace}\renewcommand{\Dm}{DM\xspace}\xspace}

\newcommand{\shs}{Shack-Hartmann sensor (SHS)\renewcommand{\shs}{SHS\xspace}\renewcommand{\shss}{SHSs\xspace}\xspace}
\newcommand{\shss}{Shack-Hartmann sensors (SHSs)\renewcommand{\shs}{SHS\xspace}\renewcommand{\shss}{SHSs\xspace}\xspace}
\newcommand{\lgs}{laser guide star (LGS)\renewcommand{\lgs}{LGS\xspace}\renewcommand{\lgss}{LGSs\xspace}\xspace}
\newcommand{\lgss}{laser guide stars (LGSs)\renewcommand{\lgs}{LGS\xspace}\renewcommand{\lgss}{LGSs\xspace}\xspace}
\newcommand{\ngs}{natural guide star (NGS)\renewcommand{\ngs}{NGS\xspace}\renewcommand{\ngss}{NGSs\xspace}\xspace}
\newcommand{\ngss}{natural guide stars (NGSs)\renewcommand{\ngs}{NGS\xspace}\renewcommand{\ngss}{NGSs\xspace}\xspace}
\newcommand{\mems}{Micro-Electro-Mechanical Systems (MEMS)\renewcommand{\mems}{MEMS\xspace}\xspace}
\newcommand{\snr}{signal to noise ratio (SNR)\renewcommand{\snr}{SNR\xspace}\xspace}
\newcommand{\moao}{multi-object \ao (MOAO)\renewcommand{\moao}{MOAO\xspace}\xspace}
\newcommand{\mcao}{multi-conjugate adaptive optics (MCAO)\renewcommand{\mcao}{MCAO\xspace}\xspace}
\newcommand{\ltao}{laser tomographic adaptive optics (LTAO)\renewcommand{\ltao}{LTAO\xspace}\xspace}
\newcommand{\cpu}{central processing unit (CPU)\renewcommand{\cpu}{CPU\xspace}\renewcommand{\cpus}{CPUs\xspace}\xspace}
\newcommand{\cpus}{central processing units (CPUs)\renewcommand{\cpu}{CPU\xspace}\renewcommand{\cpus}{CPUs\xspace}\xspace}
\newcommand{\dsps}{digital signal processors (DSPs)\renewcommand{\dsps}{DSPs\xspace}\xspace}
\newcommand{\psf}{point spread function (PSF)\renewcommand{\psf}{PSF\xspace}\renewcommand{\psfs}{PSFs\xspace}\xspace}
\newcommand{\psfs}{point spread functions (PSFs)\renewcommand{\psf}{PSF\xspace}\renewcommand{\psfs}{PSFs\xspace}\xspace}
\newcommand{\fpga}{field programmable gate array (FPGA)\renewcommand{\fpga}{FPGA\xspace}\renewcommand{\fpgas}{FPGAs\xspace}\xspace}
\newcommand{\fpgas}{field programmable gate arrays (FPGAs)\renewcommand{\fpga}{FPGA\xspace}\renewcommand{\fpgas}{FPGAs\xspace}\xspace}
\newcommand{\wpu}{wavefront processing unit (WPU)\renewcommand{\wpu}{WPU\xspace}\xspace}
\newcommand{\sor}{successive over-relaxation (SOR)\renewcommand{\sor}{SOR\xspace}\xspace}
\newcommand{\fdpcg}{Fourier domain pre-conditioned gradient (FDPCG)\renewcommand{\fdpcg}{FDPCG\xspace}\xspace}
\newcommand{\map}{maximum a-posteriori (MAP)\renewcommand{\map}{MAP\xspace}\xspace}
\newcommand{\usb}{Universal Serial Bus (USB)\renewcommand{\usb}{USB\xspace}\xspace}
\newcommand{\elt}{Extremely Large Telescope (ELT)\renewcommand{\elt}{ELT\xspace}\renewcommand{\elts}{ELTs\xspace}\xspace}
\newcommand{\elts}{Extremely Large Telescopes (ELTs)\renewcommand{\elt}{ELT\xspace}\renewcommand{\elts}{ELTs\xspace}\xspace}
\newcommand{\dugall}{Durham University generalised adaptive optics laser laboratory (DUGALL)\renewcommand{\dugall}{DUGALL\xspace}\xspace}
\newcommand{\fwhm}{full-width at half-maximum (FWHM)\renewcommand{\fwhm}{FWHM\xspace}\xspace}
\newcommand{\wht}{William Herschel Telescope (WHT)\renewcommand{\wht}{WHT\xspace}\xspace}
\newcommand{\emccd}{electron multiplying CCD (EMCCD)\renewcommand{\emccd}{EMCCD\xspace}\xspace}
\newcommand{\dasp}{Durham \ao simulation platform (DASP)\renewcommand{\dasp}{DASP\xspace}\xspace}
\newcommand{\eelt}{European \elt (E-ELT)\renewcommand{\eelt}{E-ELT\xspace}\xspace}
\newcommand{\mpi}{Message Passing Interface (MPI)\renewcommand{\mpi}{MPI\xspace}\xspace}
\newcommand{\smp}{symmetric multi-processing (SMP)\renewcommand{\smp}{SMP\xspace}\xspace}
\newcommand{\svd}{singular value decomposition (SVD)\renewcommand{\svd}{SVD\xspace}\xspace}
\newcommand{\gpu}{graphical processing unit (GPU)\renewcommand{\gpu}{GPU\xspace}\renewcommand{\gpus}{GPUs\xspace}\xspace}
\newcommand{\gpus}{graphical processing units (GPUs)\renewcommand{\gpu}{GPU\xspace}\renewcommand{\gpus}{GPUs\xspace}\xspace}
\newcommand{\fft}{fast Fourier transform (FFT)\renewcommand{\fft}{FFT\xspace}\xspace}
\newcommand{\ifu}{integral field unit (IFU)\renewcommand{\ifu}{IFU\xspace}\xspace}
\newcommand{\darc}{the Durham \ao real-time controller (DARC)\renewcommand{\darc}{DARC\xspace}\renewcommand{\Darc}{DARC\xspace}\xspace}
\newcommand{\Darc}{The Durham \ao real-time controller (DARC)\renewcommand{\darc}{DARC\xspace}\renewcommand{\Darc}{DARC\xspace}\xspace}
\newcommand{\cots}{commercial off-the-shelf (COTS)\renewcommand{\cots}{COTS\xspace}\xspace}
\newcommand{\rtcp}{real-time control pipeline (RTCP)\renewcommand{\rtcp}{RTCP\xspace}\xspace}
\newcommand{\rms}{root-mean-square (RMS)\renewcommand{\rms}{RMS\xspace}\xspace}
\newcommand{\sFPDP}{serial Front Panel Data Port (sFPDP)\renewcommand{\sFPDP}{sFPDP\xspace}\xspace}

\newcommand{\mnras}{MNRAS}

\newcommand{\aap}{A\&A}

\newcommand{\pasp}{Pub. Astron. Soc. Pacific}
\newcommand{\nar}{New Astronomy Reviews}

\begin{document}

\title{The Durham adaptive optics real-time controller}

\author{Alastair Basden,$^{1,*}$ Deli Geng,$^1$ Richard Myers,$^1$ and Eddy Younger$^{1}$}
\address{$^1$Department of Physics, Durham University, South Road,
  Durham DH1 3LE, UK}
\address{$^*$Corresponding author: a.g.basden@durham.ac.uk}

\begin{abstract}
The Durham adaptive optics real-time controller was initially a proof
of concept design for a generic adaptive optics control system.  It
has since been developed into a modern and powerful CPU based
real-time control system, capable of using hardware acceleration
(including FPGAs and GPUs), based primarily around commercial off the
shelf hardware.  It is powerful enough to be used as the real-time
controller for all currently planned 8~m class telescope adaptive
optics systems.  Here we give details of this controller and the
concepts behind it, and report on performance including latency and
jitter, which is less than 10~$\mu$s for small adaptive optics
systems.

\end{abstract}
\maketitle

\section{Introduction}
\Ao is a technology widely used in optical and infra-red astronomy,
and almost all large current and planned science telescopes have an \ao system.  A large
number of results have been obtained using \ao systems which would
otherwise be impossible for seeing-limited (uncorrected)
observations\cite{2004A&A...417L..21G,2005ApJ...625.1004M}.  New \ao
techniques are being studied for novel applications such as wide-field
high resolution imaging \cite{2004SPIE.5490..236M} and extra-solar
planet finding \cite{2004ASPC..321...39M}.

When starlight passes through the Earth's atmosphere, random
perturbations are introduced which distort the wavefronts from the
astronomical source in a time varying fashion \cite{tatarski}.  It is
then no longer possible to form a diffraction limited image from these
distorted wavefronts, and the effective resolution of a telescope is
reduced.  By sensing the form of the wavefront using a \wfs as
described by Roddier \cite{roddier}, and then rapidly applying
corrective measures to one or more deformable mirrors, it is possible
to compensate for some of the perturbations, and hence improve the
image quality and resolution of the telescope.  The \wfs and
deformable mirror together form part of an \ao system.

A real-time control system is used to interpret the \wfs signals and
compute the commands that are to be sent to the
\dm\cite{1997SPIE.3126..269D,2006PASP..118..297W,2008SPIE.7015E..95T,2006NewAR..49..618S}.
This is a computationally intensive task as new \dm commands are
computed from \wfs signals typically one thousand times per second,
and the shape of the \dm must be adjusted in real-time, before the
atmospheric turbulence has changed significantly, requiring a control
system that is able to operate with minimal latency.
It has long been thought that \cpus are not suitable for real-time
control of \ao systems because of poor performance and large jitter.
Previous systems have typically been comprised of complex arrangements
of \dsps and \fpgas, requiring extensive development time and quick
obsolescence due to the fast changing nature of new hardware.
However, this is no longer the case, and here we present a modern \cpu
based solution.

\Darc is a highly configurable \ao control platform designed for the
control of current and as-yet-unknown \ao systems.  It has been
developed from a proof of concept system into a capable controller
based around a modern architecture, running on \cots hardware.  It is
primarily a \cpu based system, with the ability to use additional
hardware acceleration where available.  The reasons for developing it
were two-fold; to have a powerful system capable of controlling high
order laboratory experiments at Durham, and to have a control system
capable of controlling a multiple \lgs \moao on-sky demonstrator \ignore{the
  CANARY \cite{canary}} instrument \cite{canary} on the \wht from 2010
onwards, giving on-sky usage of the control system.  The technology
demonstrated in this experiment is likely to impact the design of
future instruments.  \ignore{ CANARY is a \moao technology
  demonstrator which aims to prove the concept of using \lgss with
  open-loop \dm control.  The technologies demonstrated by CANARY
  should then be used by \eelt instruments.  }

Here, we give an overview of \darc in \S2, including key
components, algorithms implemented and hardware used.  In \S3 we
describe the performance of the system, and conclusions are made in
\S4.

\section{DARC overview}
\begin{figure}
\includegraphics[width=6cm]{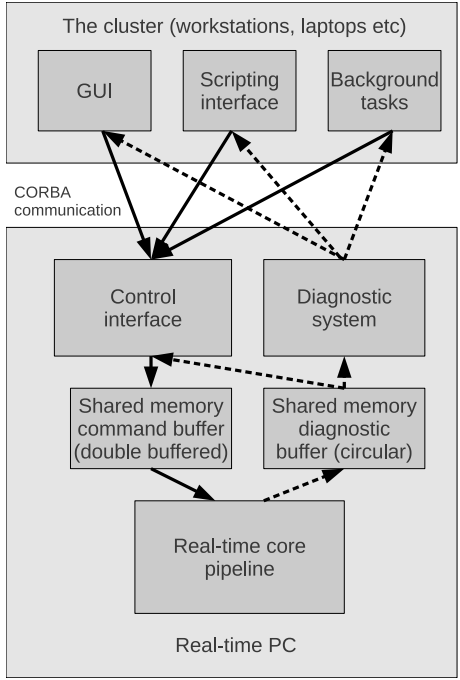}
\caption{An overview of the DARC system, showing the key components
  and interactions between them.  Solid lines show control flow and
  dashed lines show data (telemetry) flow.}
\label{fig:darcoverview}
\end{figure}

\Darc is comprised of several key components: The \rtcp, a control
interface, a diagnostic system, a graphical and scripting interface
and background tasks.  The \rtcp does the bulk of the
work, taking \wfs camera data and computing the control vectors to be
sent to deformable mirrors.  The control interface is responsible for
allowing the user to update and control the \rtcp, for example
changing reference images.  The diagnostic system is an optional
component, recommended for large systems, and is responsible for
taking output produced by the \rtcp, logging it and distributing it to
clients as requested.  The graphical and scripting interfaces provide
easy ways for a user to alter the state of the system via the control
interface.  An overview of the components of \darc are shown in
Fig.~\ref{fig:darcoverview}. 

\subsection{The real-time control pipeline}
The \rtcp is a software executable running on real-time (or standard)
Linux.  It is multi-threaded with the number of threads being user
controllable.  The user specifies the priorities of each of these
threads and can restrict them to certain processors (the \cpu
affinity), and make such changes while the \rtcp is running.  When
running on multi-core machines (currently, up to 16 cores are readily
available off-the-shelf) this gives the user flexibility to optimise a
given \ao system for given computational hardware.  The \rtcp also has
the ability to make use of the processing power found in \gpus for
wavefront reconstruction, if suitable hardware is available.

Diagnostic data is written to shared memory circular buffers and
includes raw and calibrated pixels, slope measurements, wavefront
estimates, \dm commands and flux measurements.  The \rtcp cannot be
delayed by the writing of diagnostic data, meaning that performance is
not affected by network glitches or errors in diagnostic reading
sub-systems.  

\subsubsection{Interfaces to the RTCP}
The data input to the \rtcp can either be in the form of raw camera
images, calibrated camera images or wavefront slope measurements.  An
optional \fpga front-end has been developed at Durham for the ESO
SPARTA AO system \cite{sparta}, called the \wpu, which calibrates \wfs
images and computes wavefront slopes.  This \wpu front-end can
optionally be used with the \rtcp and has very low (sub-microsecond)
latency and virtually no jitter, being deterministic.  It should be
noted that this \wpu front-end performs weighted
centre-of-gravity slope measurements; if other slope measurement
algorithms are required, they must be performed by the \rtcp.

The interface for handling data input into the \rtcp is based on
shared object libraries, which can be changed (or indeed, written)
while the \rtcp is running.  Such interfaces currently include a
\sFPDP \cite{sfpdp} interface (used with low-light-level \emccd
cameras at Durham), a Gbit Ethernet interface (used with Pulnix
cameras), a socket interface, a \usb interface (used with Xenics IR
cameras) and a file interface for testing.  A library for virtually
any interface could be written as required by a user.  The user has
the ability to swap between these interfaces as required without
stopping the \rtcp, useful for example for changing from a physical
camera into a replay mode.  This is also the case for the interface
responsible for sending the \dm commands, and currently implemented
interfaces include one for sending actuator demands to a figure sensor
over \sFPDP, one to control the \dm combination that we have at Durham
directly, and one for sending actuator demands over a socket.  Such
flexibility means that \darc can be used with practically any \ao
system without code changes to the core real-time system (and without
the subsequent debugging that this would entail), with the user
writing input and output libraries appropriate for their system.

There is also an interface for handling wavefront reconstruction,
allowing the wavefront reconstruction algorithm to be changed (and
even developed) on-the-fly without halting the \rtcp.  Currently
implemented reconstruction interfaces include a Kalman filter, a
standard matrix-vector multiplication integrator and a matrix-vector
based algorithm for open-loop wavefront control.  Additionally, a \gpu
based matrix-vector multiplication interface is also available, and
will be presented in a separate paper.  We also intend to implement
iterative and Fourier based reconstruction algorithms using this
interface.

\subsubsection{RTCP processing strategy}
The processing of data is carried out using what we term a horizontal
processing strategy where many threads perform multiple algorithms,
with each thread performing the same operations as other threads.
Data is processed by the \rtcp on a sub-aperture by sub-aperture
basis, with a thread requesting work, and being assigned one (or more)
sub-apertures to process (as soon as enough pixels have arrived for
this sub-aperture) up to the stage of computing the influence
that this sub-aperture has on the \dm (if possible, dependent on
reconstruction algorithm) before returning and requesting another
sub-aperture to process.  This allows the processing to commence as
soon as enough camera pixels have arrived for a given sub-aperture,
before the whole camera frame is ready, greatly reducing the time
between the last camera pixel being read out and the \dm commands
being ready (the latency), an important parameter for an \ao control
system.  Once the last sub-aperture has been processed the partial \dm
commands computed by each thread are amalgamated and post-processed
(including clipping) by a post-processing thread and sent to the \dm.
Whilst this post-processing is being carried out, the other threads
are able to start processing the next \wfs camera frame if it is
available.  All sub-apertures of a given frame must have been
processed before the next frame is commenced.  

\begin{figure}
\includegraphics[width=8cm]{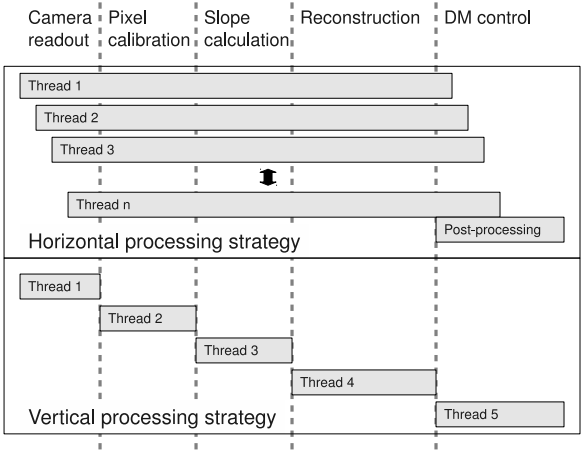}
\caption{A figure demonstrating the horizontal and vertical processing
  strategies.  With a horizontal strategy all threads perform the same
operations allowing the processing load to be balanced between
available processors.  With a vertical strategy, the work performed by
threads is unequal, load balancing becomes harder, and greater latency
is introduced by the need to communicate between each stage.}
\label{fig:processing}
\end{figure}

A more conventional processing strategy is the vertical processing
strategy where discrete tasks are assigned to different threads, or at
least, performed one at a time rather than being broken up into pieces
(e.g.\ a thread or process for image calibration, a thread for slope
computation and a thread for wavefront reconstruction).  Here,
synchronisation between stages occurs either after each stage has
completed (greatly increasing latency since all stages must wait until
all the pixel data has arrived, and then until all previous stages
have completed), or after each sub-aperture
has completed its stage.  This however greatly increases the amount of
synchronisation required between stages (using for example Mutexes or
locks) which in turn adds to the latency.  Fig.~\ref{fig:processing}
demonstrates these strategies.

A horizontal processing strategy allows far better \cpu load
optimisation than a vertical processing strategy since each thread is
assigned the same amount of work, which will allow a higher \cpu
utilisation while there is processing to be carried out, reducing the
total time taken.  Additionally, processing of a sub-aperture is
carried out from start to finish by one thread, reducing the need for
thread synchronisation that a vertical processing strategy would
require (where each processing stage needs to be synchronised with the
others).  

The ability to compute partial \dm commands (i.e.\ the influence of a
given sub-aperture on the final \dm command vector) is dependent on
the wavefront reconstruction algorithm used.  With most iterative and
Fourier based wavefront reconstruction algorithms, the computation of
partial \dm commands is not possible.  These algorithms require all of
the slope measurements to be present before the computation of the \dm
command vector commences, meaning that the wavefront
reconstruction algorithm only starts after all slope
measurements have been computed.  This has the disadvantage that a greater
amount of post processing is required after all the wavefront sensor
pixels have been received by the \rtcp, and so can result in a greater
latency than algorithms that allow the computation of partial \dm commands.

With a matrix-vector implementation (where the \dm command is updated
with the dot product of a control matrix with a vector of wavefront
slope measurements) the computation of partial \dm commands is
possible, since the column of the matrix corresponding to a given
sub-aperture is multiplied by the slope measured from this
sub-aperture (a scalar--vector multiplication) to give the influence
that this sub-aperture has on the final \dm command.  This computation
can commence as soon as the slope for this sub-aperture has been
computed, which in turn is computed as soon as enough pixels for
this sub-aperture have arrived from the wavefront sensor.  Camera
readout is relatively slow: A fast camera may take 0.1~$\mu$s to read
out each pixel, so the time to read out a sub-aperture can be large
compared with the time to compute the wavefront slope of this
sub-aperture.  The final \dm command is then computed by summing
together all the partial \dm commands.

The reconstructor interface of the \rtcp is flexible enough
to allow both algorithms that compute partial \dm commands, and
algorithms that cannot compute partial \dm commands to be used,
allowing different reconstruction algorithms to be plugged in and out
of the \rtcp.

One drawback of the horizontal processing strategy is that it is not
as easy to separate processing tasks out onto different hardware as it
would be with a vertical processing strategy, for example by using one
PC to perform image calibration, another PC to perform wavefront
reconstruction, and another to perform mirror control.  However,
feedback between reconstruction and control algorithms is often a
requirement making these unlikely to be separated.  Additionally,
image calibration is optionally performed by an \fpga front-end, such as the
\wpu developed at Durham, making a separate calibration processing
task unnecessary.  Furthermore, as we demonstrate in this paper, by
using a horizontal processing system like \darc on current top-end
processors, it is unnecessary to separate these tasks for likely \ao
systems on current 8~m class telescopes, since modern processors are
powerful enough to perform the required task.

\subsubsection{Algorithms overview}
Standard image processing algorithms such as background and noise
removal, flat fielding, thresholding and pixel weighting are
carried out by the \rtcp to calibrate the raw wavefront sensor images
(when not using the \wpu front-end).  

The wavefront slopes across each sub-aperture are computed using a
centre of gravity (standard or weighted) centroiding algorithm or a
correlation centroiding algorithm \cite{centalgo}.

The \rtcp uses one of a number of wavefront reconstruction algorithms
to compute the commands to be sent to the \dms.  Least-squares
or minimum variance reconstruction are implemented, with several
control laws including Kalman filtering \cite{kalman} being
available.  The \rtcp can be used in closed-loop (where the \wfss are
sensitive to changes on the \dm) or open-loop (where the \wfss are
placed before the \dm in the optical path, and so do not measure
changes applied to it) operation.

Slope linearisation, essential for getting best performance with
open-loop systems, is optional.  Here, a look-up table
between computed wavefront slope (non-linear due to the pixellated
nature of the detectors) and actual wavefront
slope is provided, and so computed slopes are linearised using this
look-up table.  A separate look-up table is provided for each
sub-aperture and \wfs, allowing for different non-linearities to be
compensated, for example with elongated spots due to a laser guide
star.  The linearisation calibration should be performed using a tip-tilt
mirror, gradually tilting the wavefront across the \wfs, and recording
slope measurements as this is carried out.  The non-linearity is
usually small enough that closed-loop systems are not adversely
affected since slope measurements are minimised by the \dm, and arises
from the discrete pixellated nature of the detectors and is also
caused by optical effects (imperfect lenslets for example).

An adaptive windowing algorithm is also available with two operating
modes.  With a global mode, the mean spot motion is tracked and the
pixels assigned to each sub-aperture follow this mean spot motion.
With a per-sub-aperture adaptive windowing mode, the pixels assigned
to each sub-aperture track the motion of the spot in this sub-aperture
rather than being fixed.  Adaptive windowing is a useful feature for
open-loop control, where spot motions may be large, especially for
\lgss.

\subsubsection{Optional figure sensing operation}
\begin{figure}
\includegraphics[width=8cm]{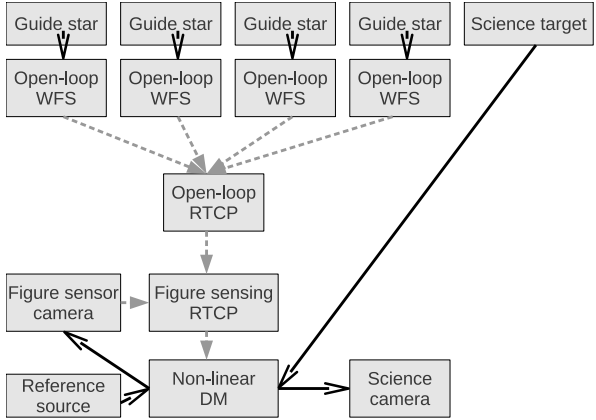}
\caption{A schematic diagram showing a typical configuration for an
  open-loop AO system used with a figure sensor.  Solid black arrows with
  unfilled arrow heads show optical paths, while dashed grey arrows with
  filled arrow heads show electrical paths.  The open-loop RTCP
  receives wavefront data from open-loop wavefront sensors.  The
  mirror demands (an ideal mirror is assumed) are passed to the figure
  sensing RTCP, which combines these with the measured shape of the
  mirror (obtained using the figure sensor wavefront sensor), to set
  the mirror to the desired shape.}
\label{fig:figuresensor}
\end{figure}

The \rtcp may optionally accept an asynchronous \dm command input in
addition to the \wfs input, enabling it to be used as a figure sensor
alongside an open-loop \ao system, allowing a non-linear \dm to be
observed and controlled so that the actual shape can be tweaked until
it matches the requested shape.  This allows for open-loop operation
with a non-linear \dm (wavefront sensors do not measure the shapes
applied to the \dm, so do not know its true shape).  In this case one
instance of the \rtcp would compute desired \dm commands (assuming a
linear \dm) using the open-loop (on-sky) \wfs measurements.  This
\rtcp can assume that it has a perfect, linear \dm attached.  These
commands are then sent to a second instance of the \rtcp (on the same,
or different hardware) which would have image input from a figure
sensing \wfs that is in closed loop control with the non-linear
physical \dm, operating at a higher frame-rate than the on-sky \wfss.
This second \rtcp rapidly measures the physical \dm surface and
adjusts the commands sent to the \dm until the \dm surface has reached
the shape that the first \rtcp has requested, as shown in
Fig.~\ref{fig:figuresensor}.  It would do this by reading the figure
sensor \wfs at a rate faster than the main open-loop \rtcp.  The
wavefront reconstructed from the figure sensor \wfs data then provides
the current actual \dm shape (typically using a least-squares
integrator control law) which is compared with the desired \dm shape,
and the difference applied to the \dm.  The figure sensing control
loop typically runs at between two and five times faster than the main
open-loop \rtcp.  This allows open-loop operation with \dms that have
poor characteristics such as large hysteresis and a non-linear
response.  Fig.~\ref{fig:figuresensor} shows a typical configuration
for an open-loop \ao system with a figure sensor.  It should be noted,
that the on-sky, open-loop \rtcp does not receive any feedback from
the \dm or from the figure sensing \rtcp.

\subsection{The control interface}
The control interface runs on the same hardware as the \rtcp, and
accesses shared memory to update the \rtcp parameters.  The parameters
are double-buffered and so the control interface does not delay the
\rtcp while parameters are updated.  The control interface exposes a
CORBA object \cite{corba} which is then used by clients to perform any
required operations.  CORBA, the Common Object Request Broker
Architecture, is an standard architecture and infrastructure that
allow programs from almost any computer, operating system and
programming language to interoperate.

The control interface is controlled remotely using scripts or a
graphical interface, to perform operations such as computing system
interaction (response) matrices, calculating and setting background
maps, and fine tuning algorithms.  It can also be used to obtain
diagnostic data.  

\subsection{The diagnostic system}
\Darc implements a philosophy of separate diagnostic streams for all
diagnostic data (images, slope measurements, etc.).  These streams are
turned on or off as required, depending on demand and the ability of
the system to cope (considering network bandwidth, processing power,
etc.).  Additionally, streams can be independently decimated with data
for every $N^{th}$ frame being sent.  \Darc offers two different
sources for diagnostic streams.  For most systems, a diagnostic server will be
implemented, typically running on different computational hardware.
This server receives diagnostic streams from the \rtcp via a process
that takes data from the \rtcp shared memory circular buffers and then
distributes this data to clients as requested, removing the processing
load from the \rtcp machine.  For small scale \ao systems and
laboratory tests, diagnostic data can be sent using the control
interface avoiding the complexity of the diagnostic server, allowing
the control system to be set up quickly with minimal hardware in a
laboratory environment.

Clients subscribe to diagnostic streams using a CORBA interface and are
then sent the data when it is ready.  Diagnostic streams can be switched
on or off, and decimated in three separate places: At the \rtcp, in
the diagnostic server, and on a client by client basis.  This allows
fine control of how much data should be sent, for example every 10
frames of raw pixels for logging (by the diagnostic server), but every
20 frames of raw pixels sent to client A, and every 25 frames of raw
pixels send to client B.  This ability allows the performance of \darc
to be fine tuned to available hardware for a given system size.

\subsection{The graphical interface and scripting}
A graphical interface is used to control the real-time system
remotely, display current status and parameters, and view diagnostic
streams.  This tool is generic and is used with systems with any
number of wavefront sensors and deformable mirrors.  More than one
instance of the graphical interface may be run simultaneously.  This
tool uses the CORBA objects of the control interface and the
diagnostic system to realise the necessary control.

Scripting is also carried out in a similar way, and a user script may be
written in any language for which CORBA is available.  There are also
some command line tools for performing simple operations with \darc
(for example setting parameters and obtaining diagnostic data).

\subsection{Background tasks}
Background tasks, such as control matrix optimisation using turbulence
profiling data, and optimisation of system performance are be performed
remotely, using data from the diagnostic system, and updating the
\rtcp using the control interface.  Such operations are dependent on
\ao system type and configuration and so should be developed for each
system on which \darc is deployed, rather than being part of \darc,
which is designed to be generic.  

\section{System performance}
\begin{figure}
\includegraphics[width=8cm]{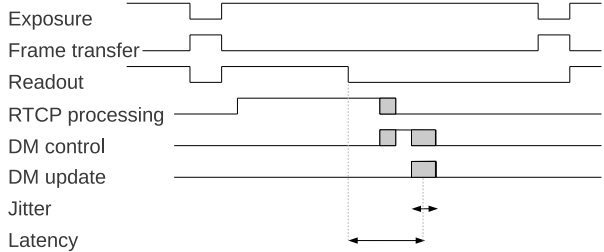}
\caption{A schematic timing diagram for an AO system with a frame
  transfer CCD.  Jitter in signals is shown by the grey areas (where
  time of transition may vary), which is introduced by the
  non-deterministic processes of RTCP processing and DM control.  }
\label{fig:timingdiagram}
\end{figure}

When evaluating the system performance of an \ao control system, there
are a number of performance parameters that should be measured.
Fig.~\ref{fig:timingdiagram} shows a schematic timing diagram for an
\ao system, assuming a frame transfer CCD.  Here, \rtcp processing
commences once enough pixels have been received, and is concurrent
with CCD readout.  Latency is defined as the time from end of readout
to average \dm update time, while jitter is the variation in \dm
update time, the stability of the system.  Low jitter is crucial for
an \ao system, as there is little point having a very low average
latency (e.g.\ 1~$\mu$s) if it sometimes takes a long time to process
a frame (e.g.\ 1~s) because the scientific image quality would drop
during this delay.  The latency is an important parameter because it
affects the mean age of the \wfs image data at the time when the \dm
is updated, equal to $t_a = t_e/2 + t_r + t_l$ where $t_e$ is exposure
time, $t_r$ is readout time, $t_l$ is the latency and $t_a$ is the
mean data age.  A larger mean data age means a less accurate \dm
correction since the atmosphere will have had greater time to evolve,
which will result in reduced performance, as the \dm shape will no
longer be matched to the evolved atmospheric disturbance.  Minimal
latency is essential for best performance.  The maximum \rtcp
frame-rate is the maximum rate at which the system can process camera
images (though usually this will be limited by the camera).  It should
be noted that in a \rtcp with a vertical processing strategy (see
Fig.~\ref{fig:processing}) the latency can be long even when the
maximum frame-rate is high, since multiple frames may be at different
stages of processing simultaneously.  For example frame one could be
at the processing stage where actuator commands are being produced,
while frame two is at the slope measurement stage, frame three is at
the image calibration stage and frame four is being read from the
\wfs.  Conversely, with a horizontal processing strategy such as we
have implemented with \darc, the latency will be approximately the
inverse of the maximum frame-rate (assuming the case where this is not
limited by cameras), since the processing of one frame has to complete
before processing of the next frame commences (post-processing may add
some latency).  Hence by reducing the latency of the \rtcp the maximum
possible frame rate is increased.  This then equates into hardware
savings because less hardware can be used to run a given \ao system at
the required frame-rate.  Less hardware also provides a system that is
simpler to manage, and more reliable (fewer components to break).

We have tested the performance of \darc using a dual quad-core Intel
Xeon based computer and a quad quad-core AMD Opteron based computer
(both 2008 specification with 5500 series Xeon processors at 2.26~GHz
and 8378 series Opteron processors at 2.4~GHz respectively) and both
give similar performance for small and medium sized systems.  It
should be noted that recent Intel chipsets perform a house-keeping
task approximately once per second which needs switching off (using a
Linux kernel patch) to avoid occasional jitter when these tasks
perform.  The \rtcp results presented here are using a standard Linux
(non-real-time) kernel on the quad quad-core Opteron based system.
The \rtcp will also run with a real-time kernel without modification,
further reducing jitter, though we do not present such results here
because the camera interface drivers for which we are making these
measurements are not compatible with a real-time kernel.

\subsection{Oscilloscope latency and jitter measurements}
We have measured the latency and jitter of \darc using a digital
oscilloscope triggered on the camera exposure trigger (start of
readout), looking at the variation of when changes in the voltages
sent to the \dm drive electronics occur.  Similar measurements were
also recorded by the \rtcp (the time taken to compute a frame) and
compared.  The oscilloscope is set with infinite persistence, meaning
that newly triggered signals are overlaid on existing signals,
allowing a record of signal transition times to build up, giving a
measure of the variation over which these transitions occur.  We
transition the \dm actuators from low to high or vice-versa every
camera frame and so the \dm signal is at half the frequency of the
camera signal (high on one camera frame, low the next, and so on {\it
  ad infinitum}).  Combined with the infinite persistence, this means
that the \dm transitions from low to high and from high to low are
shown overlain.  The mean distance of the set of actuator transitions
from the camera trigger pulse gives the mean latency of the system
once the camera readout time has been subtracted.  The width of the
set of actuator transitions gives a measure of the jitter.  The
latency is assumed to have a Gaussian distribution so the
oscilloscope trace built up over a long period of time (infinite
signal persistence) will approximately display the mean latency $\pm 3
\sigma$ where $\sigma$ is the \rms jitter.  Therefore we record the
\rms jitter as $\frac{1}{6}$ of the measured long time average
actuator transition width.

\subsubsection{Single camera system}
Our bench system used for the tests presented here is comprised of a
single $128\times128$ pixel \emccd camera with $7\times7$
sub-apertures and a 52 actuator \dm with a weighted centre-of-gravity
slope calculation algorithm, a least-squares matrix-vector based
wavefront reconstruction algorithm and an integrator control law.  It
should be noted that in this system, the \rtcp sends actuator demands
to a separate PC running a figure sensor which then places these
demands onto the mirror (the \dm control time in
Fig.~\ref{fig:timingdiagram}).  In these tests, actuator values are
just passed unaltered to the \dm, however latency and jitter
associated with an additional sFPDP transfer and additional PC are
included in these results.  A figure sensor should be an integral part
of any open-loop \ao system (such as \moao) design when using \dms
with a non-linear response, being the only way we have of knowing what
shape the \dm is actually taking, and so it was included in these
performance tests.

\ignore{
\begin{figure}
\includegraphics[width=8cm]{091118_125411BasdenF4.eps}
\caption{(Color online) An oscilloscope trace showing camera trigger pulse (yellow)
  and the transition of DM actuators (blue), with a latency of
  $620\pm8$~$\mu$s (measured from the distance between the blue and
  yellow vertical transitions with the uncertainty given by the width
  of the blue vertical transition) .  Pixels are arriving at the
  maximum sFPDP rate.  This trace has infinite persistence, meaning
  that many signals are shown overlaid allowing a picture of the
  average transition time to build up.  The DM signal is at half the
  frequency of the camera signal, since we transition the actuators
  every camera frame (high one frame, low the next, etc.), and
  transitions from low to high and from high to low are both overlaid.  }
\label{fig:scopejitter}
\end{figure}
}
\ignore{
In Fig.~\ref{fig:scopejitter} the start of camera readout signal is
shown by the yellow trace while the blue trace shows a \dm actuator
command which switches low to high (and vice versa) every frame (and
hence is half the camera frequency).  The vertical blue lines near the
middle of the figure represent multiple instances of when the actuator
commands are applied with a peak to valley difference of about
50~$\mu$s (bounded by the yellow bars), and a corresponding \rms
jitter of about 8~$\mu$s (assuming a Gaussian distribution with six
standard deviations visible).  
}

The mean time between start of camera readout (trigger pulse) and
setting of \dm actuators is measured using the oscilloscope to be
about 620~$\mu$s when the total camera readout time is 400~$\mu$s (the
maximum pixel rate limited by the data transfer rate to the \rtcp).  A
\rms jitter of 8~$\mu$s is measured (a peak to valley jitter of
50~$\mu$s) giving a latency measurement of $220\pm8$~$\mu$s.  After
recording actuator transitions for about a minute we typically see
that there are three outlying \dm transitions during this time, the
largest of which adds an additional 65~$\mu$s latency (which may be
due to the figure sensor).  A 15 minute recording over a larger
oscilloscope timebase to check for large occasional latency variations
shows no actuator transitions away from the main set.  Therefore large
infrequent changes in latency are not occurring.  It should be noted
that this is not an oscilloscope selection effect, and that if we
manually perform a high priority intensive task on the computer
(running at a higher priority than the \rtcp) then single lines become
visible at larger latencies.  \ignore{ The camera pixel data in this
  case is fake data generated by an \fpga and is arriving at close to
  the maximum rate of the \sFPDP fibre interface.  }

\ignore{
\begin{figure}
\includegraphics[width=8cm]{091118_130832BasdenF5.eps}
\caption{(Color online) An oscilloscope trace showing camera trigger pulse (yellow)
  and the transition of DM actuators (blue).  Pixels are arriving at
  the maximum sFPDP rate.  The vertical yellow bars are set to the
  position of the trigger and the position of the actuator
  transition.  This trace has infinite persistence,
  meaning that many signals are overlaid, allowing a picture of the
  average transition time to build up.  }
\label{fig:scopejitter2}
\end{figure}
}

When the camera is running as typically used on-sky (i.e.\ with good
image quality) it has a readout time of about 2000~$\mu$s, so the
last pixel arrives at the \rtcp about 2000~$\mu$s after the trigger
pulse.  Using the oscilloscope to measure the actuator transition
times relative to the trigger pulse, we find that the \dm commands are
sent $2100$~$\mu$s after the trigger pulse, meaning there is about
100~$\mu$s between the last camera pixel being received by the \rtcp
and the \dm actuators being set.  The latency in this case is
therefore $100\pm6$~$\mu$s since the oscilloscope measured latency has
a peak to valley difference of 36~$\mu$s.  This is reduced compared
with the previous measurement because the \rtcp has more time for
calculation while waiting for camera pixels to arrive.

\begin{figure}
\includegraphics[width=8cm]{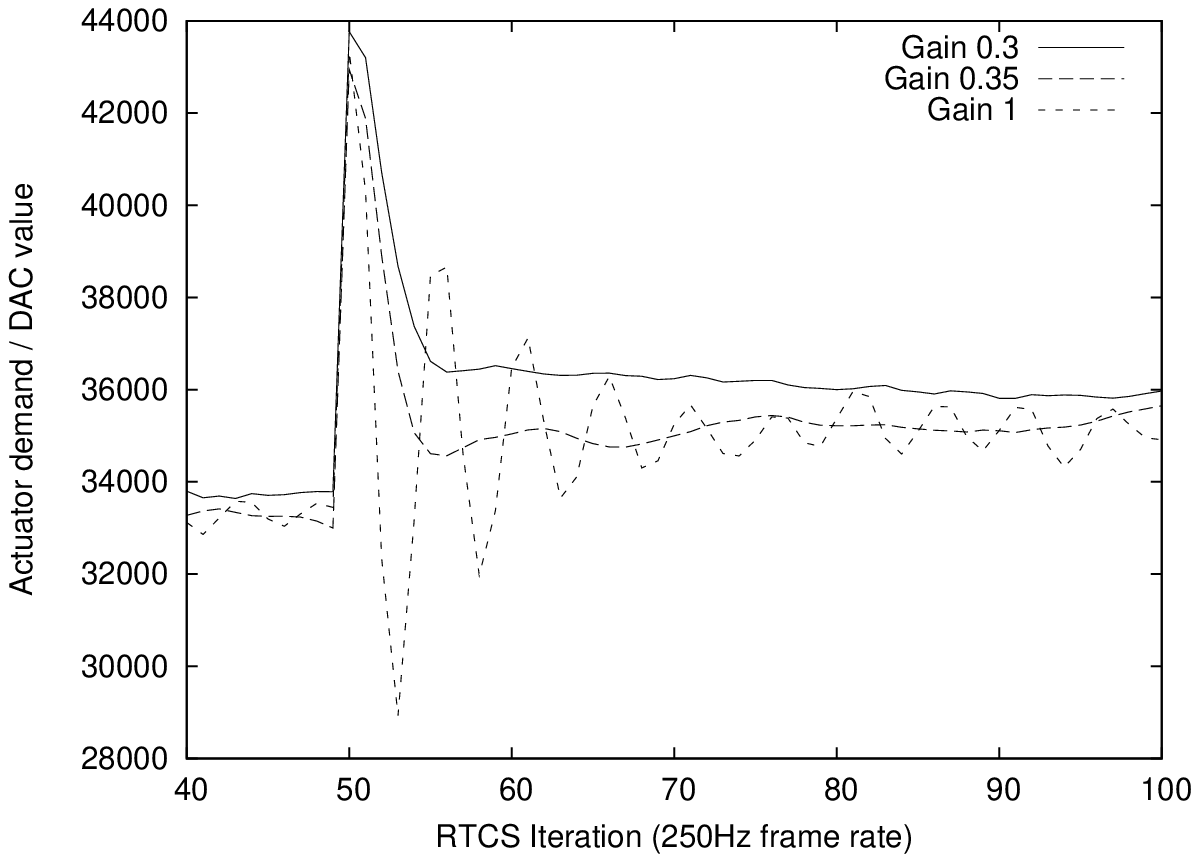}
\includegraphics{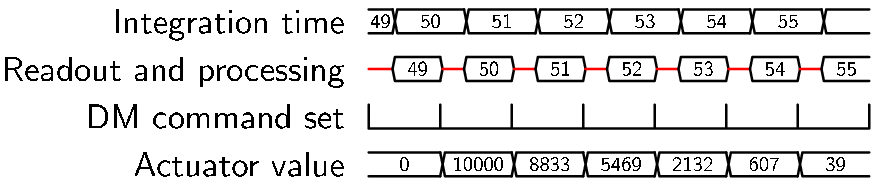}
\caption{(a) Actuator response to an applied perturbation as a
  function of time.  The plot shows the command applied to the DM
  actuator, after a step change of 10000 DAC values was inserted at
  iteration 50.  (b) (Color online) A schematic timing diagram demonstrating the
  expected number of iterations before the perturbation has been
  corrected, for a gain of 0.35, with readout and processing
  take two-thirds of the exposure time.}
\label{fig:response}
\end{figure}

We have measured the actuator response time of the \rtcp by operating
the system in closed loop mode, and then applying an offset voltage to
one actuator, while recording the commands sent to the \dm.
Fig.~\ref{fig:response}(a) shows the \ao system response time while
operating at different integrator gains.  In this figure, an offset
voltage is applied at iteration 50, and the \rtcp subsequently
corrects this offset.  With a unity gain (with which the system can
close the loop stably without blowing up, demonstrating the low
latency performance), a fast actuator response is seen with correction
being applied in only two iterations.  Overshoot, or ringing occurs,
due to the necessary delay between an image being integrated on the
detector and read out (taking finite time), and the shape being
applied on the \dm.  With a gain of 0.3, the system is under-damped,
while a gain of 0.35 shows critical damping, taking about four
iterations for the loop compensation to be fully applied, as expected
from analysis of the timing diagram in Fig.~\ref{fig:response}(b).
The actuator values given in this timing diagram are calculated by
assuming that the \rtcp will compute the actuator value to be a time
weighted mean of the two states that the actuator is in during the
integration time (taking two-thirds of the time in one state, and
one-third of the time in the next state).  The integrator then
multiplies this by the gain, and adds to the previous integrator
state.  The actuator value is seen to have returned to near zero after
only a short number of iterations.

\subsubsection{Figure sensor latency}
The latency and jitter of the figure sensor PC are included in these
measurements, as mentioned previously.  We have measured separately
the latency and jitter for the figure sensor PC alone and obtain
$70\pm5$~$\mu$s latency with some outliers that double this value
(recorded over approximately 30000 transitions).  To make these
measurements, a trigger pulse is generated in an \fpga, which then
also implements a fake camera interface with 52 pixels (since there
are 52 \dm actuators) which are passed straight to the figure sensor
and interpreted as a \dm command vector.

\ignore{
However,
Fig.~\ref{fig:scopedmc}\ignore{091120_123649_DMC.png} shows the
measurement of latency and jitter for the figure sensor PC alone,
corresponding to a $70\pm5$~$\mu$s latency with some outliers that can
double this value (recorded over approximately 30000 transitions).
\begin{figure}
\includegraphics[width=8cm]{091120_123649_DMCBasdenF8.eps}
\caption{(Color online) An oscilloscope trace showing the pulse upon which commands
  are sent to the figure sensor (yellow) and the transition of DM
  actuators (blue) giving a latency of $70\pm5$~$\mu$s.  The vertical
  yellow bars bound the transition of the DM actuators, and hence give
  a measure of the jitter.  The trace has infinite persistence, so
  multiple instances of triggering and signal recording are overlaid.}
\label{fig:scopedmc}
\end{figure}
}

\subsubsection{Software measured jitter}
\begin{figure}
\includegraphics[width=8cm]{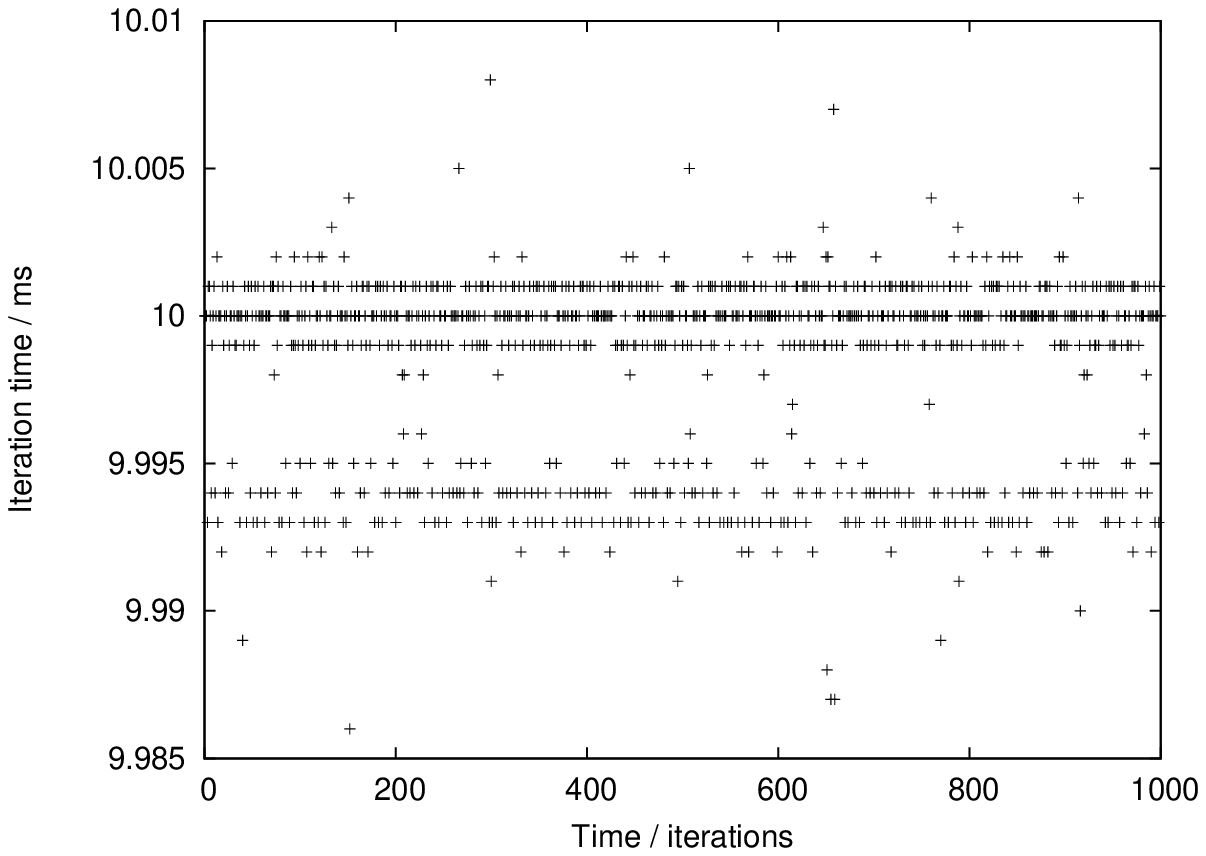}
\includegraphics[width=8cm]{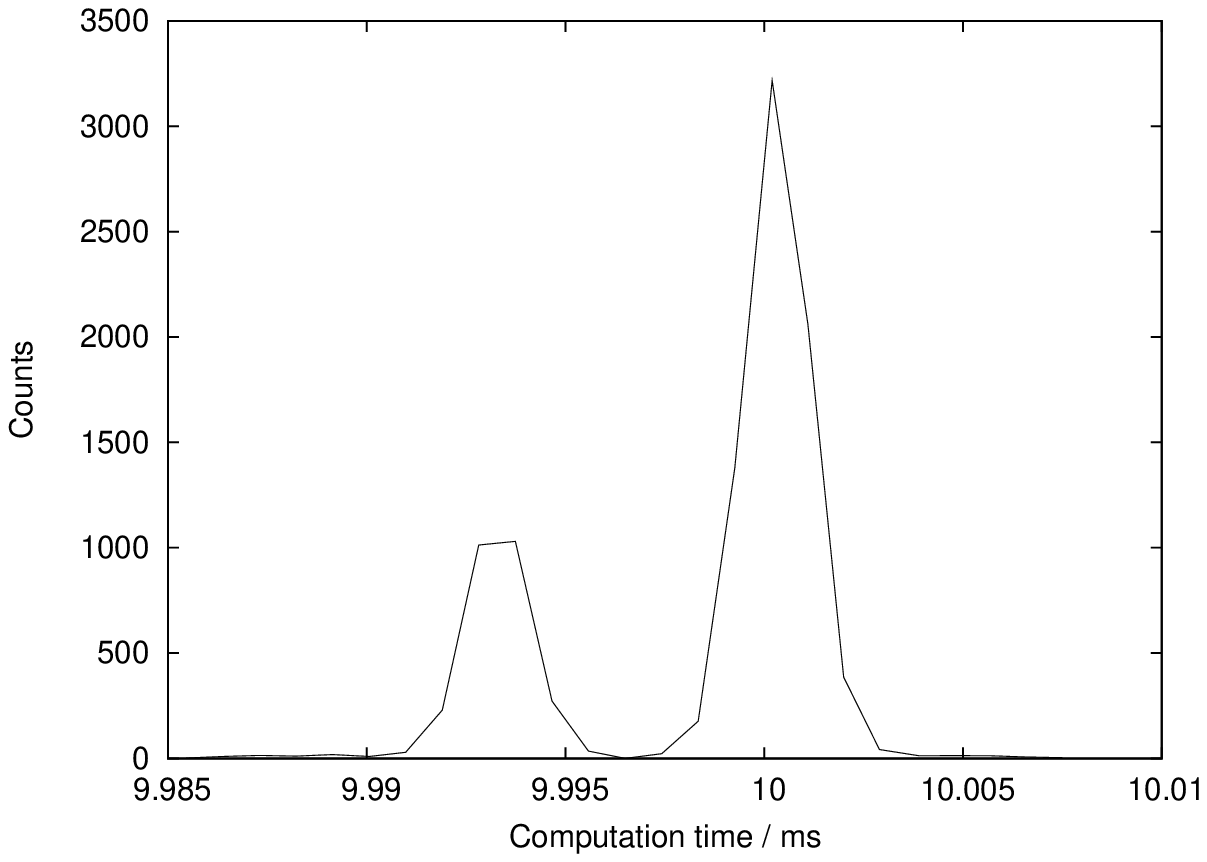}

\caption{(a)~Frame iteration time generated by the real-time control
  pipeline over 1000 iterations.  (b)~A histogram of the frame
  computation time (10000 iterations).  Camera frequency is 100~Hz (10~ms frame time).}
\label{fig:softjitter}
\end{figure}

The \rtcp is able to record the time taken to process frames using the
PC hardware clock.  Fig.~\ref{fig:softjitter}\ignore{use
  timing091120-155413.fits - real camera} shows timings for 10000
frames, with a mean value of just under 10~ms (the camera frame rate),
a worst case measurement of 10~$\mu$s above this and a \rms of
3~$\mu$s.  This is in good agreement with the oscilloscope estimated
value of 8~$\mu$s which also includes the figure sensor jitter of
about 5~$\mu$s (note, these should be added in quadrature) and is only
estimated from the width of the oscilloscope trace.  The histogram
shown in Fig.~\ref{fig:softjitter}(b) shows that the latency
distribution is non-Gaussian, having a double peak, which would cause
an oscilloscope estimated jitter value to be a slight overestimation.
Our oscilloscope measurements also give some evidence for this double
peak though this is not conclusive due to the crude nature of this
measurement (``transition'' or ``not transition'' but no record of
number of transitions at a given time).  We are unsure of the reason
for this double peak, which may be a result of the software timer
resolution.  Alternatively it could be due to scheduling in the Linux
kernel or could be a true feature of the \rtcp.

\subsection{Other AO system configurations}
We have measured the maximum frame-rate that \darc running on our
available hardware is capable of handling for a number of different
\ao system configurations.  Software recorded jitter measurements are
given, though since we do not have suitable cameras for these cases,
the oscilloscope measurements are not available.

The cases we consider are shown in table~\ref{table:setups} for a
simple scaling of a single guide-star, single \dm system.  In these
cases the frame time is equal to the time taken to process an image
and compute the actuator demands, but not to receive the image (since
we don't have appropriate cameras) or to send the actuator commands
(since we don't have appropriate mirrors).

\begin{table}
\begin{tabularx}{\linewidth}{XXXXX}
\hline
Sub-apertures & Image size & Actuators & Frame time / $\mu$s & With \wpu
front-end / $\mu$s \\ \hline
$8\times8$ & $128\times128$ & 52       & $149\pm12$ & $20\pm2$ \\
$16\times16$ & $256\times256$ & 208    & $519\pm22$ & $35\pm3$ \\
$32\times32$ & $512\times512$ & 832    & $1517\pm37$ & $83\pm3$ \\
$64\times64$ & $1024\times1024$ & 3328 & $9160\pm85$ & $355\pm3$ \\ \hline
\end{tabularx}
\caption{A table showing measured frame computation time for different AO
  system configurations.  The results with \wpu front-end assume that
  slope measurements are received from an FPGA and so only wavefront
  reconstruction is performed using the RTCP.}
\label{table:setups}
\end{table}
\ignore{Data taken from: timings091120-*.fits where * is 135924,
  140248, 143144, 143330, 143616, 143737, 144341, 144612}

Measurements are given for cases which include the Shack-Hartmann
wavefront sensor image calibration and slope calculation time, and
also for cases where the calibration and slope calculation is carried
out by the \wpu front-end (which is \fpga based and has about
0.5~$\mu$s latency, so is negligible), allowing the \rtcp to be used
for wavefront reconstruction, \dm command vector calculation and
house-keeping tasks.

It should be noted that these measurements are real, made with the
\rtcp, not extrapolations, and show that \darc has the ability to
operate a classical $32\times32$ sub-aperture \ao system at over
500~Hz using a single PC, or to operate a $64\times64$ sub-aperture
\ao system at over 2~kHz using a single PC and a \wpu front-end.  This
would give the performance of a planet finding instrument on a 8~m
class telescope and demonstrates the ability of \darc when compared
with the computational hardware traditionally required by high order
\ao systems\cite{sparta} involving large numbers of hardware based
processing boards, subsystems and complex communications protocols.

\subsection{On-sky tests}
The technology demonstrator instrument on the \wht with which the
real-time controller is used \ignore{CANARY \moao instrument
  \cite{canary}} has three separate phases.  At phase A, from
September 2010 onwards, there are four on-sky \wfss, three of which
are used for open-loop tomography (wavefront reconstruction) and do
not see the \dm, while the fourth is used as a truth sensor behind the
\dm, to measure the corrected wavefront, though is not used in the
reconstruction process.  There is a tip-tilt mirror, and a 52 channel
\dm neither of which are visible to the on-sky \wfss.  Phase B
(beginning in 2011) will introduce four laser guide star \wfss and
phase C introduces an additional, high order (1024 actuator) \dm, all
of which are to be controlled by \darc.

We have investigated the \rtcp performance for phase A when computing
the open-loop \dm demands for the three open-loop \wfs configuration,
each with $128\times128$ pixels and $7\times7$ sub-apertures.  A mean
\rtcp computed latency of $245\pm11$~$\mu$s for the total processing
time without the \wpu front-end is achieved, well within the required
specification of $1000\pm100$~$\mu$s.

\begin{figure}
\includegraphics[width=8cm]{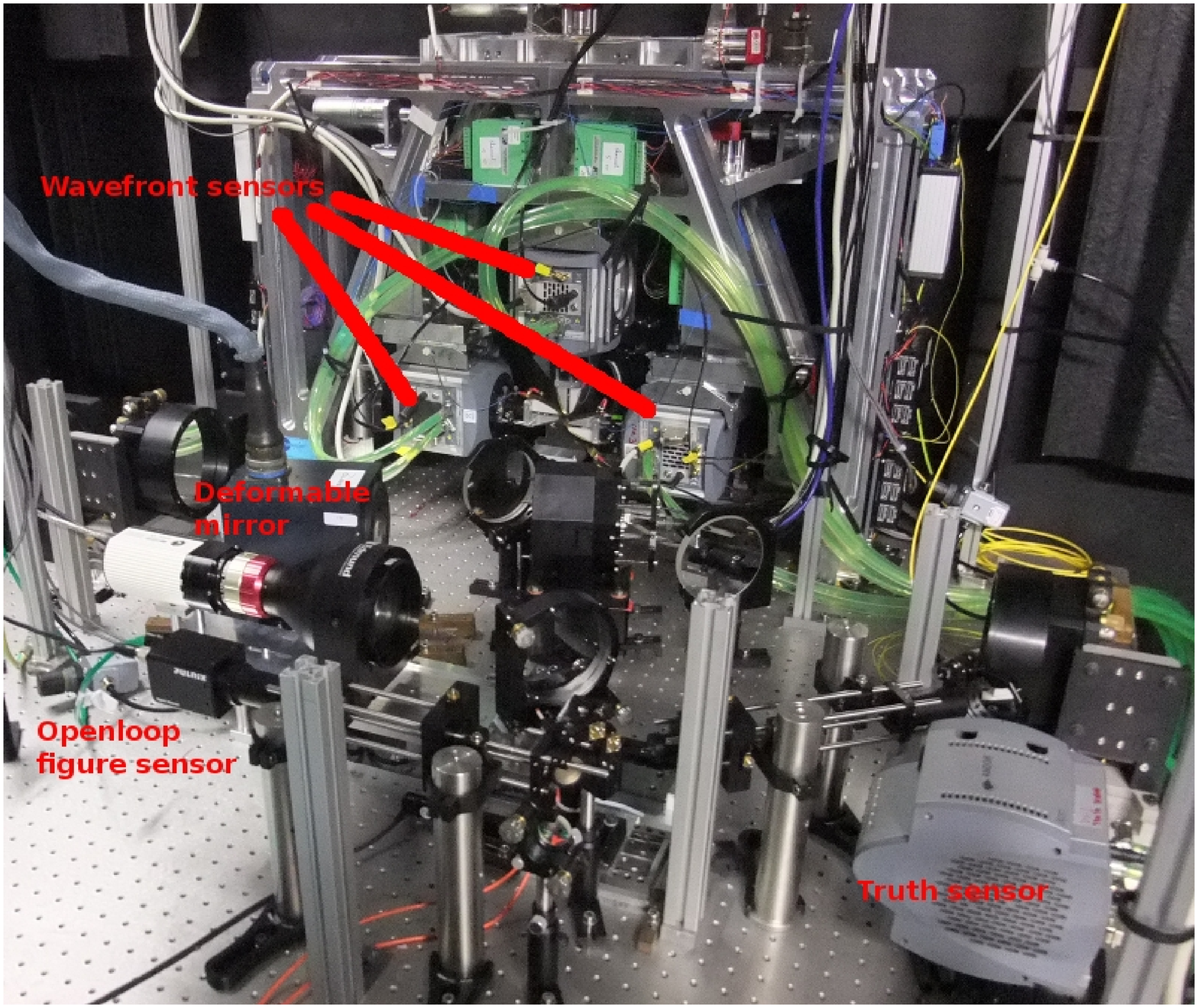}
\caption{(Color online) A picture of the technology demonstrator test bench at the
  WHT, with the AO system components clearly visible.}
\label{fig:bench}
\end{figure}

Fig.~\ref{fig:bench} shows the optical bench used for the phase A
configuration, with \wfss, truth sensor, and \dm 
visible.  A transfer function for this system gives a latency of
800~$\mu$s between last pixel received and \dm command applied when
operating in closed loop, processing wavefront data from all four
\wfss and using the truth sensor measurements to close the loop.  This
measurement includes the \dm response time of about 500~$\mu$s.  

\begin{figure}
\includegraphics[width=4cm]{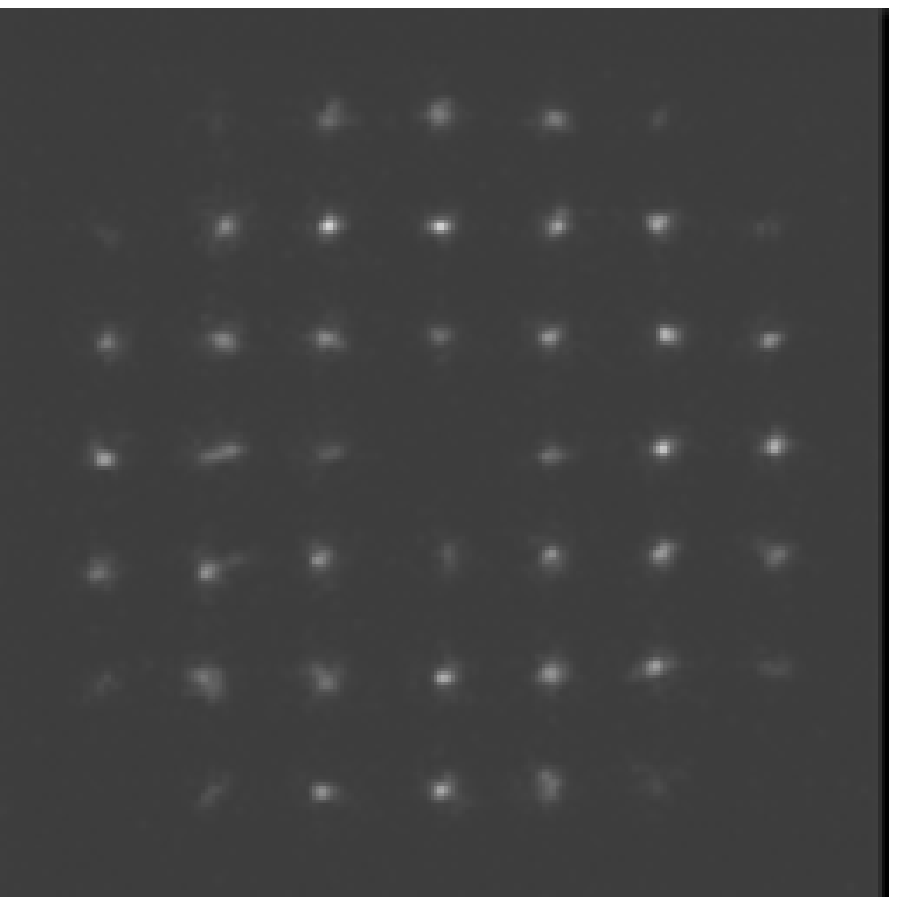}
\includegraphics[width=4cm]{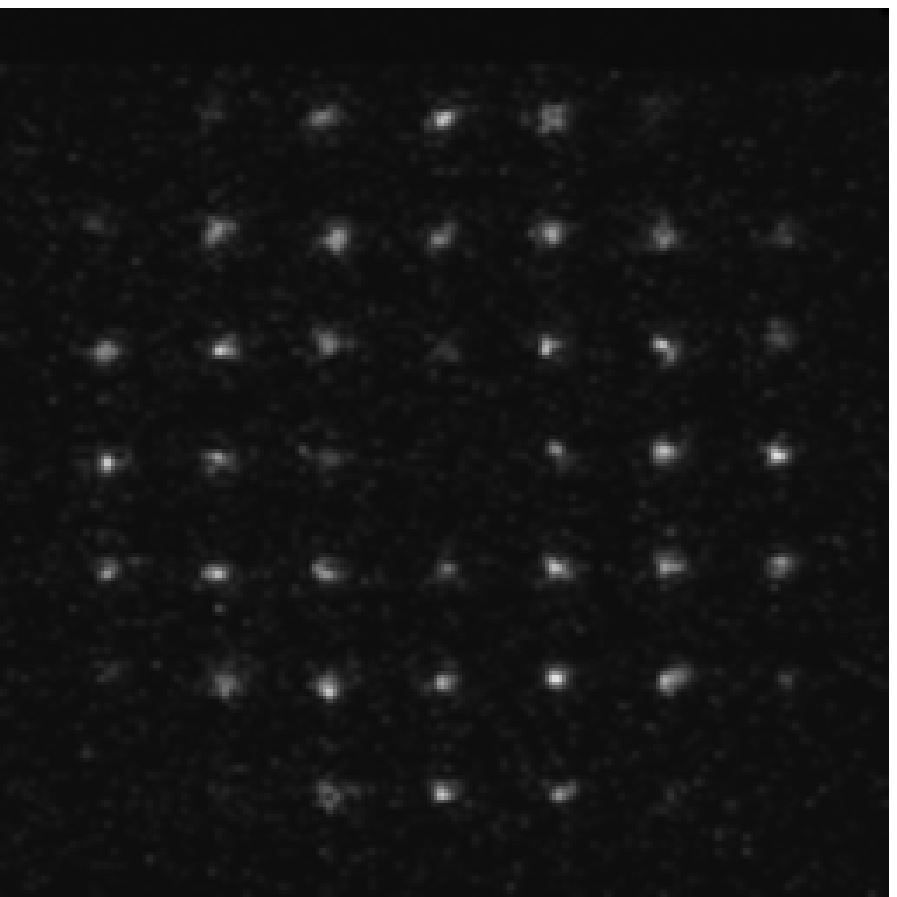}
\includegraphics[width=4cm]{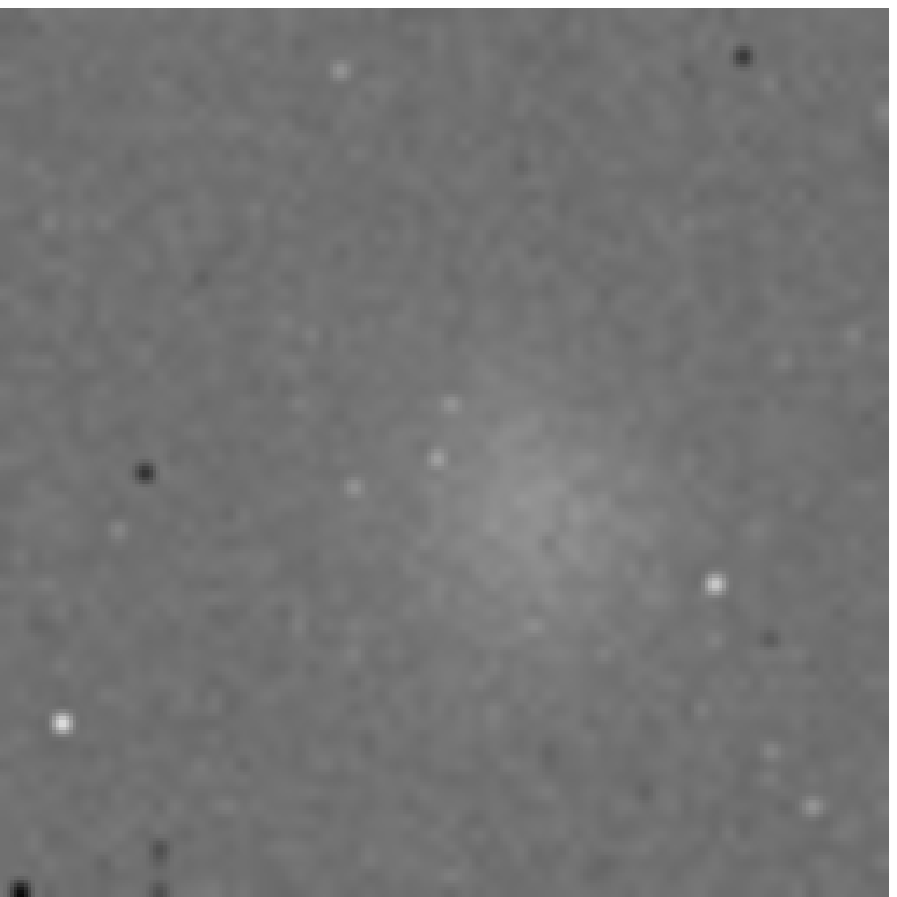}
\includegraphics[width=4cm]{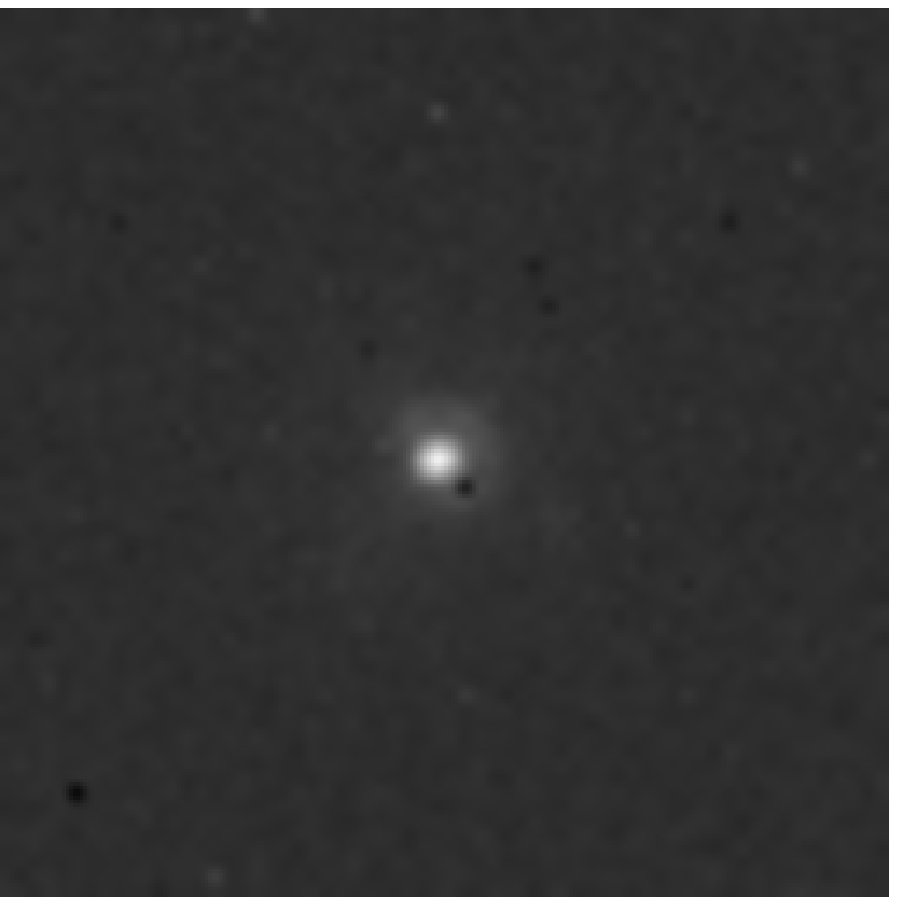}
\caption{(a) Sample open-loop wavefront sensor measurements.  (b)
  Sample truth sensor measurements.  (c) An uncorrected H-band image.  (d) A
  MOAO corrected H-band image.}
\label{fig:wfsdata}
\end{figure}

Sample (optical) \wfs data is shown in Fig.~\ref{fig:wfsdata}(a) from
one of the open-loop \wfss imaging a 8.7 Mag$_v$ star.  Corresponding
truth sensor measurements (for an 11 Mag$_v$ star) when the loop was
engaged, taken by the \wfs behind the \dm (which is tomographically
controlled by the open-loop \wfss), are shown in
Fig.~\ref{fig:wfsdata}(b).  Uncorrected and corrected H-band science
images for this setup are shown in Fig.~\ref{fig:wfsdata}(c) and (d),
with a corresponding Strehl ratio about 26\% for the corrected image.

\subsection{Maximum frame rates}
For \darc, the maximum frame rates that can be achieved for a given
system are the inverse of the latency due to the horizontal processing
strategy (assuming a fast enough camera).  For example,
table~\ref{table:setups} shows that a frame-rate of greater than 2~kHz
is achieved for a $64\times64$ sub-aperture system when using the \wpu
front-end, which would be suitable for a 8~m class telescope with a
planet finder instrument.  It should be noted that there is little
point in operating an \ao system with a frame time significantly less
than that of the \dm response time, since in this case, the \dm
convergence curve will show errors increasing before the overall error
is reduced.  However, since some \dms can have an extremely fast
response, and since \darc is designed as a generic \ao real-time
control system, it is important that the performance of \darc is maximised.

\ignore{
\section{Multi-rate wavefront sensors}
Although it is not currently possible to use the \rtcp with multi-rate
or asynchronous wavefront sensors, the inclusion of this ability would
not be difficult, and we are intending to carry out this development
in the near future.  An \ao system with multiple wavefront sensors
will typically view a number of guide stars of different
brightnesses.  There is then a trade-off to be made: Should the system
be run at a fast rate such that the fainter sources are very noisy, or
should the system be run at a slower rate to reduce the noise, but
also increase the system latency, reducing performance.  In this section, we
describe the algorithms that will allow the \rtcp to be used with
asynchronous wavefront sensors.
}

\section{Conclusion}
We have given details of the Durham \ao real-time controller, and an
overview of some of the features of this system and the algorithms it
provides.  This is a \cpu based system with an optional \fpga based
\wpu front-end for performing pixel calibration and slope calculation,
and the ability to optionally perform wavefront reconstruction using a
\gpu.  This real-time control system is generic, not targeted at any particular \ao
system, and yet powerful.  
The performance of this system has been investigated for a number of
different \ao system configurations, and the latency and jitter
measured.  This system meets the requirements of the \moao technology
demonstrator instrument on the William Herschel Telescope.  The
latency of the \rtcp when used with a 3 \wfs system and a 52 actuator
\dm (demonstration phase A) is measured to be about 245~$\mu$s with a
\rms jitter of 11~$\mu$s, well within the required specification.

When using an \fpga front-end, the latency of the \rtcp when used with
a extreme \ao system with a $64\times64$ sub-aperture wavefront
sensor, driving 3328 mirror actuators is 355~$\mu$s with a \rms jitter
of 3~$\mu$s, allowing the \rtcp to exceed a 2~kHz frame-rate typically
required for most planet-finding instruments on 8-10~m class
telescopes.

The low latency and jitter provided by \darc demonstrates that current
\cpu technology is suitable as a candidate for use with moderate sized
\ao systems, and that expensive and complex custom hardware designs
are no longer necessary for such systems.  Such low latency is
necessary to maximise the performance of \ao systems, allowing changes
to be applied to \dms before the atmospheric perturbations have had
time to evolve significantly.

\section*{Acknowledgements}
This work is funded by the STFC.  The authors would also like to thank the
team who used \darc so well on-sky.


\begin{thebibliography}{10}
\newcommand{\enquote}[1]{``#1''}

\bibitem{2004A&A...417L..21G}
E.~{Gendron}, A.~{Coustenis}, P.~{Drossart}, M.~{Combes}, M.~{Hirtzig},
  F.~{Lacombe}, D.~{Rouan}, C.~{Collin}, S.~{Pau}, A.-M. {Lagrange},
  D.~{Mouillet}, P.~{Rabou}, T.~{Fusco}, and G.~{Zins}, \enquote{{VLT/NACO
  adaptive optics imaging of Titan},} \aap \textbf{417}, L21--L24 (2004).

\bibitem{2005ApJ...625.1004M}
E.~{Masciadri}, R.~{Mundt}, T.~{Henning}, C.~{Alvarez}, and D.~{Barrado y
  Navascu{\' e}s}, \enquote{{A Search for Hot Massive Extrasolar Planets around
  Nearby Young Stars with the Adaptive Optics System NACO},} ApJ \textbf{625},
  1004--1018 (2005).

\bibitem{2004SPIE.5490..236M}
E.~{Marchetti}, R.~{Brast}, B.~{Delabre}, R.~{Donaldson}, E.~{Fedrigo},
  C.~{Frank}, N.~N. {Hubin}, J.~{Kolb}, M.~{Le Louarn}, J.~{Lizon},
  S.~{Oberti}, R.~{Reiss}, J.~{Santos}, S.~{Tordo}, R.~{Ragazzoni},
  C.~{Arcidiacono}, A.~{Baruffolo}, E.~{Diolaiti}, J.~{Farinato}, and
  E.~{Vernet-Viard}, \enquote{{MAD status report},} in \enquote{Advancements in
  Adaptive Optics. Edited by Domenico B. Calia, Brent L. Ellerbroek, and
  Roberto Ragazzoni. Proceedings of the SPIE, Volume 5490, pp. 236-247
  (2004).},  (2004), pp. 236--247.

\bibitem{2004ASPC..321...39M}
D.~{Mouillet}, A.~M. {Lagrange}, J.-L. {Beuzit}, C.~{Moutou}, M.~{Saisse},
  M.~{Ferrari}, T.~{Fusco}, and A.~{Boccaletti}, \enquote{{High Contrast
  Imaging from the Ground: VLT/Planet Finder},} in \enquote{ASP Conf. Ser. 321:
  Extrasolar Planets: Today and Tomorrow,}  (2004), pp. 39--+.

\bibitem{tatarski}
V.~I. {Tatarski}, \emph{{Wavefront Propagation in a Turbulent Medium}} (Dover,
  1961).

\bibitem{roddier}
F.~{Roddier}, \emph{{Adaptive Optics in Astronomy}} (Cambridge University
  Press, 1999).

\bibitem{1997SPIE.3126..269D}
R.~G. {Dekany}, J.~K. {Wallace}, G.~{Brack}, B.~R. {Oppenheimer}, and
  D.~{Palmer}, \enquote{{Initial test results from the Palomar 200-in. adaptive
  optics system [3126-33]},} in \enquote{Society of Photo-Optical
  Instrumentation Engineers (SPIE) Conference Series,} , vol. 3126 of
  \emph{Presented at the Society of Photo-Optical Instrumentation Engineers
  (SPIE) Conference}, {R.~K.~Tyson \& R.~Q.~Fugate}, ed. (1997), vol. 3126 of
  \emph{Presented at the Society of Photo-Optical Instrumentation Engineers
  (SPIE) Conference}, pp. 269--+.

\bibitem{2006PASP..118..297W}
P.~L. {Wizinowich}, D.~{Le Mignant}, A.~H. {Bouchez}, R.~D. {Campbell},
  J.~C.~Y. {Chin}, A.~R. {Contos}, M.~A. {van Dam}, S.~K. {Hartman}, E.~M.
  {Johansson}, R.~E. {Lafon}, H.~{Lewis}, P.~J. {Stomski}, D.~M. {Summers},
  C.~G. {Brown}, P.~M. {Danforth}, C.~E. {Max}, and D.~M. {Pennington},
  \enquote{{The W. M. Keck Observatory Laser Guide Star Adaptive Optics System:
  Overview},} \pasp \textbf{118}, 297--309 (2006).

\bibitem{2008SPIE.7015E..95T}
T.~N. {Truong}, A.~H. {Bouchez}, R.~G. {Dekany}, J.~C. {Shelton}, M.~{Troy},
  J.~R. {Angione}, R.~S. {Burruss}, J.~L. {Cromer}, S.~R. {Guiwits}, and J.~E.
  {Roberts}, \enquote{{Real-time wavefront control for the PALM-3000 high order
  adaptive optics system},} in \enquote{Society of Photo-Optical
  Instrumentation Engineers (SPIE) Conference Series,} , vol. 7015 of
  \emph{Presented at the Society of Photo-Optical Instrumentation Engineers
  (SPIE) Conference} (2008), vol. 7015 of \emph{Presented at the Society of
  Photo-Optical Instrumentation Engineers (SPIE) Conference}.

\bibitem{2006NewAR..49..618S}
R.~{Stuik}, R.~{Bacon}, R.~{Conzelmann}, B.~{Delabre}, E.~{Fedrigo},
  N.~{Hubin}, M.~{Le Louarn}, and S.~{Str{\"o}bele}, \enquote{{GALACSI The
  ground layer adaptive optics system for MUSE},} \nar \textbf{49}, 618--624
  (2006).

\bibitem{canary}
R.~M. {Myers}, Z.~{Hubert}, T.~J. {Morris}, E.~{Gendron}, N.~A. {Dipper},
  A.~{Kellerer}, S.~J. {Goodsell}, G.~{Rousset}, E.~{Younger}, M.~{Marteaud},
  A.~G. {Basden}, F.~{Chemla}, C.~D. {Guzman}, T.~{Fusco}, D.~{Geng}, B.~{Le
  Roux}, M.~A. {Harrison}, A.~J. {Longmore}, L.~K. {Young}, F.~{Vidal}, and
  A.~H. {Greenaway}, \enquote{{CANARY: the on-sky NGS/LGS MOAO demonstrator for
  EAGLE},} in \enquote{Society of Photo-Optical Instrumentation Engineers
  (SPIE) Conference Series,} , vol. 7015 of \emph{Presented at the Society of
  Photo-Optical Instrumentation Engineers (SPIE) Conference} (2008), vol. 7015
  of \emph{Presented at the Society of Photo-Optical Instrumentation Engineers
  (SPIE) Conference}.

\bibitem{sparta}
E.~{Fedrigo}, R.~{Donaldson}, C.~{Soenke}, R.~{Myers}, S.~{Goodsell},
  D.~{Geng}, C.~{Saunter}, and N.~{Dipper}, \enquote{{SPARTA: the ESO standard
  platform for adaptive optics real time applications},} in \enquote{Society of
  Photo-Optical Instrumentation Engineers (SPIE) Conference Series,} , vol.
  6272 of \emph{Presented at the Society of Photo-Optical Instrumentation
  Engineers (SPIE) Conference} (2006), vol. 6272 of \emph{Presented at the
  Society of Photo-Optical Instrumentation Engineers (SPIE) Conference}.

\bibitem{sfpdp}
{ANSI VITA}, \enquote{{Serial Front Panel Data Port},} Tech. Rep. {ANSIVITA
  17.1-2003}, {VITA} (2003).

\bibitem{centalgo}
S.~J. {Thomas}, S.~{Adkins}, D.~{Gavel}, T.~{Fusco}, and V.~{Michau},
  \enquote{{Study of optimal wavefront sensing with elongated laser guide
  stars},} \mnras \textbf{387}, 173--187 (2008).

\bibitem{kalman}
C.~{Petit}, F.~{Quiros-Pacheco}, J.~{Conan}, C.~{Kulcsar}, H.~{Raynaud},
  T.~{Fusco}, and G.~{Rousset}, \enquote{{Kalman-filter-based control for
  adaptive optics},} in \enquote{Society of Photo-Optical Instrumentation
  Engineers (SPIE) Conference Series,} , vol. 5490 of \emph{Presented at the
  Society of Photo-Optical Instrumentation Engineers (SPIE) Conference},
  {D.~Bonaccini Calia, B.~L.~Ellerbroek, \& R.~Ragazzoni}, ed. (2004), vol.
  5490 of \emph{Presented at the Society of Photo-Optical Instrumentation
  Engineers (SPIE) Conference}, pp. 1414--1425.

\bibitem{corba}
http://www.omg.org/cgi-bin/doc?formal/00 10-33.pdf, \emph{{The Common Object
  Request Broker: Architecture and Specification}} (2000).

\end{thebibliography}


\end{document}